\documentclass[%
 reprint,
 amsmath,amssymb,
 aps,
]{revtex4-2}

\usepackage{graphicx}
\usepackage{dcolumn}
\usepackage{bm}
\usepackage{booktabs}

\usepackage{xcolor}

\newcommand{\la}[1]{\label{#1}}
\newcommand{\eq}[1]{\eqref{#1}}

\def\[{\left[}
\def\]{\right]}
\def\({\left(}
\def\){\right)}
\def\d{\partial}
\newcommand{\beq}{\begin{equation}}
\newcommand{\eeq}{\end{equation}}
\newcommand\beqa{\begin{eqnarray}}
\newcommand\eeqa{\end{eqnarray}}

\begin{document}

\preprint{APS/123-QED}

\title{Integrability and Conformal Bootstrap: One Dimensional Defect CFT}

\author{Andrea Cavagli\`a$^a$}
    \email{andrea.cavaglia@kcl.ac.uk}
\author{Nikolay Gromov$^{a,b}$}
    \email{nikolay.gromov@kcl.ac.uk}
\author{Julius Julius$^a$}
    \email{julius.julius@kcl.ac.uk}
\author{Michelangelo Preti$^a$}%
    \email{michelangelo.preti@kcl.ac.uk}
\affiliation{%
 $^a$ Department of Mathematics, King's College London, Strand WC2R 2LS \\
 $^b$ St.~Petersburg INP, Gatchina, 188 300, St.~Petersburg, Russia
}%

\date{\today}

\begin{abstract}
In this letter we study how the
exact non-perturbative integrability methods in $4D$ ${\cal N}=4$ Super--Yang-Mills can work efficiently together with the numerical conformal bootstrap techniques to go beyond the spectral observables and access previously  unreachable quantities such as correlation functions at finite coupling. In the setup of 1D defect CFT living on a Maldacena-Wilson line, we managed to compute with good precision a non-supersymmetric structure constant for a wide range of the `t Hooft coupling. Our result is particularly precise at strong coupling and matches well with the recent analytic results of Meneghelli and Ferrero.
\end{abstract}

\maketitle


\section{Introduction}
When it comes to non-perturbative calculations in QFTs, the number of tools is rather limited. Monte Carlo simulation  could give rather accurate results in some theories. In a smaller set of integrable QFTs one can get access to a large number of observables, but usually those theories are limited to 2D. There are a few exceptions, such as $\mathcal{N}$=4 supersymmetric Yang-Mills (SYM) in 4D, where  integrability gives access to non-perturbative physics. The most accurate non-perturbative results are limited to the spectrum of anomalous dimensions, where the Quantum Spectral Curve (QSC)~\cite{Gromov:2013pga,Gromov:2014caa} method allows one to get extremely precise  results at `t Hooft coupling $\lambda\simeq 1$ (for example $60$ digits in \cite{Gromov:2015vua} for the BFKL Pomeron intercept).

In this paper we focus on planar $\mathcal{N}$=4 SYM. As a conformal theory, for its complete solution, in addition to the dimensions of the operators, one  also needs to know the OPE coefficients or structure constants.
Even though there have been many exciting applications of integrability to the study of the structure constants~\cite{Basso:2015zoa,Cavaglia:2018lxi,Jiang:2019zig,Cavaglia:2021mft},  these results are  either limited to perturbation theory, infinitely long operators,
strongly $\gamma$-deformed (Fishnet) theories, or are not explicit enough for numerical evaluation.
\begin{figure}
{    \centering
\includegraphics[scale=0.88]{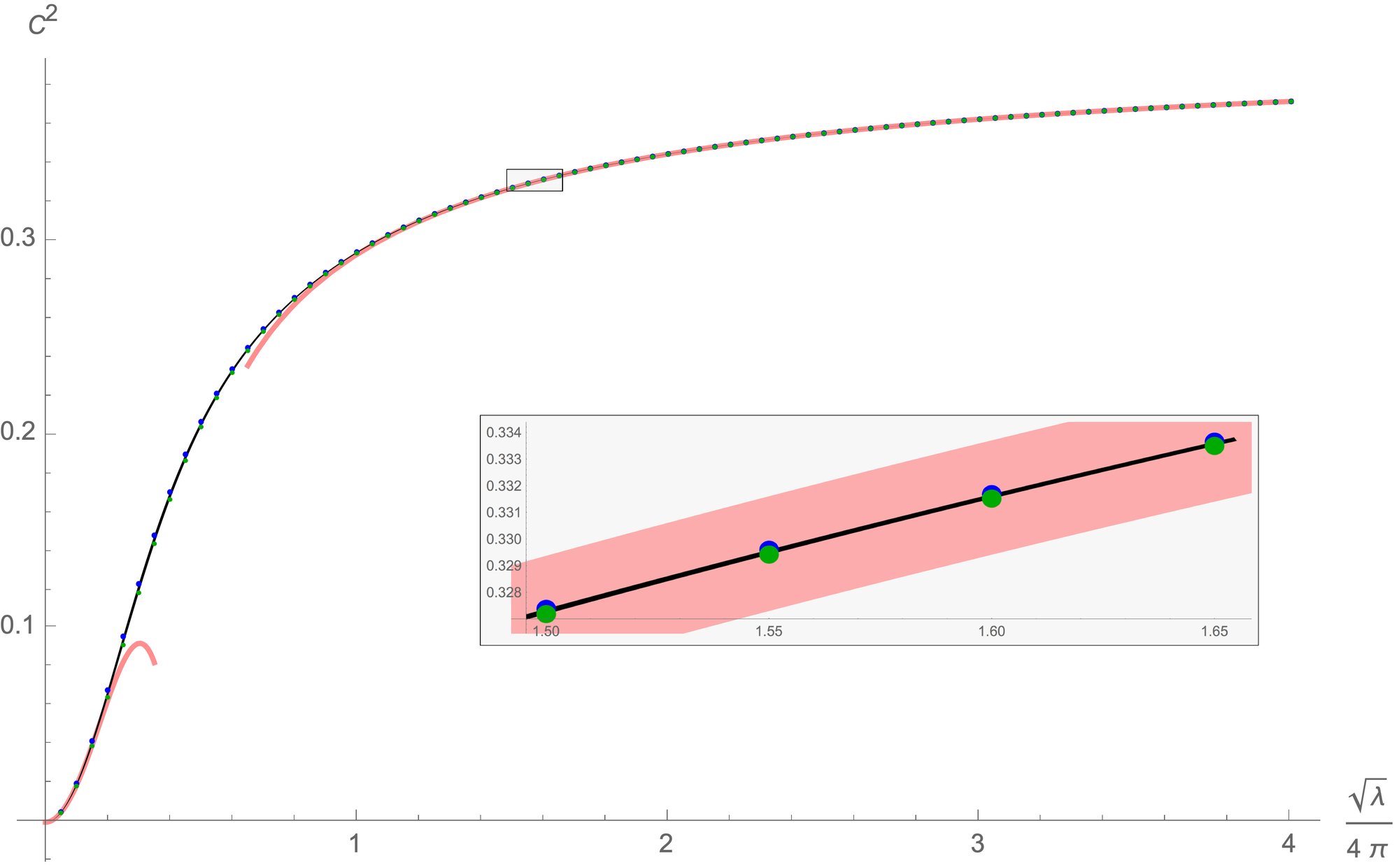}
}
    \caption{\small The OPE coefficient of two protected line-deformation operators $\Phi^i_{\perp}$ into unprotected operator $\Phi_{||}$. 
    The thickness of the black line indicates the precision. 
    The interval of values of $g=\frac{\sqrt\lambda}{4\pi}$, span a wide range in $\lambda\in(0,2526.6)$.
    It interpolates perfectly between the weak~\cite{Kiryu:2018phb,
    upcoming}
    and strong~\cite{Giombi:2017cqn,Liendo:2018ukf,Ferrero:2021bsb}
    coupling analytic results indicated by thick pink lines. We see that already at $g=1.5$ our result matches perfectly the $3$-loop result of~\cite{Ferrero:2021bsb}, obtained by a very different method. At weak coupling we compare with the preliminary result $2 g^2+\frac{4}{3} \left(\pi ^2-18\right) g^4$ of~\cite{upcoming}. 
    \label{fig:C1plot}}
\end{figure}

Another powerful non-perturbative method to study CFTs  is the 
Numerical Conformal Bootstrap (NCB) (see \cite{Simmons-Duffin:2016gjk,Poland:2018epd,Chester:2019wfx} for some reviews). It allows one to identify prohibited domains for the conformal dimensions or OPE coefficients given the symmetries of the CFT as an input.

In this paper we join together the
two powerful methods of QSC and NCB to obtain accurate values for a  nontrivial structure constant in a wide range of the coupling (see Fig.~\ref{fig:C1plot}).

Even though there is hope for 
an integrability based non-perturbative analytical 
solution of the theory, for example built upon the SoV approach~\cite{Cavaglia:2018lxi,Giombi:2018qox}, it is crucial to have an alternative numerical method which would generate high precision data for future comparison. 

\section{Setup}
Here we focus on the one-dimensional defect CFT that lives on the infinite straight 1/2-BPS Maldacena-Wilson line (MWL) in $\mathcal{N}=4$ SYM defined by \cite{Maldacena:1998im,Erickson:2000af}
\begin{align}
\label{eqn:MWLdefine}
    {\cal W} = \operatorname{Tr}W_{-\infty}^{+\infty} \equiv \operatorname{Tr}\operatorname{P}\exp\int_{-\infty}^{+\infty}dt(i\, A_t + \Phi_{||})\;.
\end{align}
$\Phi_{||}$ is one of the six real scalars. We denote the remaining $5$ scalars  by $\Phi^{i}_{\perp}$.
The MWL preserves an ${\rm OSp}(4^{*}|4)$ subgroup of the symmetry of ${\cal N}=4$ SYM. It includes: an ${\rm SO}(5)_R$ subgroup of R-symmetry, the 1D conformal group ${\rm SO}(1,2)$ and the ${\rm SO}(3)$
group of rotations in the subspace orthogonal to the line \cite{Liendo:2016ymz}.  
We mention that a similar line defect CFT setup was also considered in the 3D ABJM theory, see \cite{Bianchi:2017ozk,Bianchi:2018scb}, and the related conformal bootstrap was studied in  \cite{Bianchi:2020hsz}, while the integrability side is  not fully developed yet~\footnote{The QSC is known for the spectrum of local operators in ABJM theory \cite{Cavaglia:2014exa,Bombardelli:2017vhk}, but its version for the defect CFT is still unknown.}.

Correlation functions in the defect CFT are the expectation values of the Wilson line with insertions of local operators \cite{Drukker:2006xg,Drukker:2012de,Correa:2012at,Bonini:2015fng,Giombi:2017cqn,Cooke:2017qgm,Giombi:2018qox,Giombi:2018hsx}:
\begin{multline}
    \left\langle\left\langle {\cal O}_{1}\left(t_{1}\right) {\cal O}_{2}\left(t_{2}\right) \cdots {\cal O}_{n}\left(t_{n}\right)\right\rangle\right\rangle  \\ 
    \equiv
    \langle\operatorname{Tr}W_{-\infty}^{t_1}{\cal O}_1(t_1)W_{t_1}^{t_2}{\cal O}_2(t_2)\ldots {\cal O}_n(t_n) W^{+\infty}_{t_n}\rangle
    \;,
\end{multline}
where $W_{t_i}^{t_f}$ is a segment of the line~\eqref{eqn:MWLdefine}. 

An important role is played by so-called line-deformation BPS multiplets ${\cal B}_n$. For example the simplest
${\cal B}_1$ super-multiplet contains $\Phi^{i}_{\perp}$ as its top component. As a consequence, the dimension of this operator is protected and equal to $1$. The simplest non-protected operator is $\Phi_{||}$ \cite{Liendo:2016ymz,Cooke:2017qgm,Liendo:2018ukf,Grabner:2020nis}.

In addition to the conformal symmetry, the integrability of ${\cal N}=4$ SYM also implies that integrability governs the spectrum of this defect CFT as we describe in the next section.

\section{QSC and the spectrum}
\begin{figure}[t]
    \centering
\includegraphics[scale=0.88]{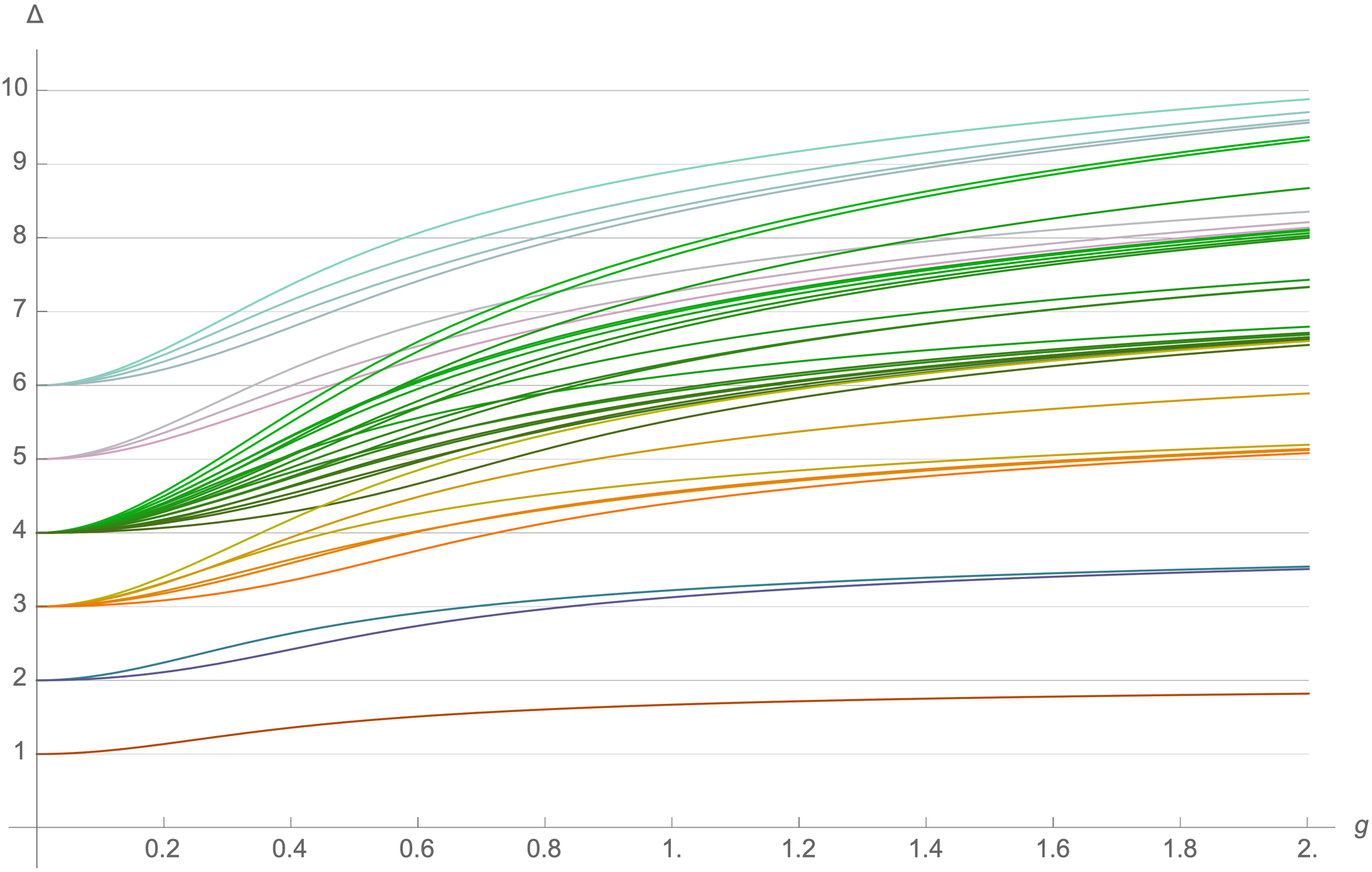}
    \caption{ Dimensions of $35$ states  computed with high precision with the  QSC using an improved method of \cite{upcoming}. For several lowest lying  states we have data in a wider range $g\in[0,4]$. The oscillator content and one-loop anomalous dimensions of the states are given in table~\ref{tab:states} of the supplementary material.}
    \label{fig:spectrum}
\end{figure}
The two point functions of conformal primary operators are  controlled by the conformal dimensions $\langle\langle O_A(t_1)O_B(t_2)\rangle\rangle \propto \delta_{AB}|t_1-t_2|^{-2\Delta_A}$.
Initially, TBA equations describing the spectrum of some operators were written in~\cite{Drukker:2012de,Correa:2012hh}, which then were transformed into a QSC form more suitable for practical calculations in~\cite{Gromov:2015dfa}, but for a long time it was not clear if integrability could  also compute dimensions of ``neutral" operators such as $\Phi_{\parallel}$.
In \cite{Grabner:2020nis} it was shown that the straight-line limit of the QSC of \cite{Gromov:2015dfa} captures those operators too. It is currently expected that all operators appearing in this defect CFT can be studied in a similar way. In particular we managed to find solutions of the QSC for all operators with nontrivial anomalous dimensions appearing in the OPE of $\Phi^i_{\perp}\times\Phi^i_{\perp}$ with bare dimensions $\Delta_0=1,2,3$ and partially for $\Delta_0=5$ and $6$ (for $\Delta_0=4$ further work is needed to confirm the state counting), which constitute  $35$ states in total (see Figure \ref{fig:spectrum}).

\paragraph{QSC.} For the details of the QSC constructions we refer to recent reviews~\cite{Gromov:2017blm,Kazakov:2018ugh,Levkovich-Maslyuk:2019awk}. Let us briefly summarise the  construction: there are $4+4$ functions ${\bf p}_a$ and ${\bf q}_i$ of one complex variable $u$, which are related by some finite difference relations. All these functions have a quadratic branch cut starting at $\pm 2g$ in the complex plane of $u$. Introducing $x(u)=\frac{u+\sqrt{u-2g}\sqrt{u+2g}}{2g}$ one can represent ${\bf p}_a$ functions as an expansion ${\bf p}_a=\sum_n \frac{c_{a,n}}{x^n}$. After that one solves for ${\bf q}_i$, which satisfy a finite difference equation in $u$ with  coefficients built out of ${\bf p}$'s. To select physical states, one has to impose the ``gluing condition" ${\bf q}_k(u\pm i0)= M_{k}^j {\bf q}_j(-u\pm i0)$ for $u\in[-2g,2g]$ where $M^i_i=1,\;M_3^2=-M_4^1=\alpha\sinh(2\pi u)$ 
and other elements $0$, which fixes the coefficients $c_{a,n}$ to a discrete set of values. The dimensions $\Delta$ can be read off the large $u$ asymptotics: ${\bf q}_i(u) \sim u^{n_i + \Delta}$ (for some integers $n_i$).

In order to find this large number of states, we have to resolve the main technical problem preventing us from getting good starting points for the numerical algorithm.
We reformulate the optimisation problem for finding the coefficients in ${\bf p}_a$ into the search for the zeros of a vector function for Fourier modes of the gluing condition, which we then solve with the Newton method.
Details of the construction will be published elsewhere~\cite{upcoming}. This method is significantly more stable at weak coupling and allows us to use the perturbative analytical solutions of the QSC (for example \cite{Marboe:2014gma,Gromov:2015vua,Marboe:2018ugv}) as starting points for the numerical algorithm efficiently even for highly excited states.

\paragraph{States.} The non-protected states appearing in the OPE of two ${\cal B}_1$ multiplets should have all quantum numbers zero (except $\Delta$)~\cite{Liendo:2018ukf}. 
In general the mixing problem is quite complicated and there are no simple closed sectors. However, at one loop, mixing is limited. For example the scalars $\Phi_{||}$ and $\Phi_{\perp}^i \Phi_{\perp}^i$ form a closed sector and the mixing matrix is known explicitly~\cite{Correa:2018fgz}. We notice that the spectrum of this mixing matrix  coincides exactly with the spectrum of a one-loop ${\rm PSU}(2,2|4)$ effective Bethe ansatz for a particular choice of Bethe roots numbers. 
It is convenient to use the  oscillator notation~\cite{Bars:1982ep,Gunaydin:1984fk,Beisert:2003jj,Beisert:2004ry},
which translates into roots  numbers easily (see e.g.~\cite{Marboe:2017dmb}). For those states we find $n_{{\bf f}_i}=
\{\Delta_0+1,\Delta_0+1, \Delta_0, \Delta_0\}$
and $n_{{\bf a}_i}=n_{{\bf b}_i}=0$. 
Plugging those numbers into the BAE solver of~\cite{Marboe:2017dmb}, and selecting states satisfying a parity in $u$ (accompanied with a flip of the Dynkin diagram), we reproduce the  spectrum of the mixing matrix of~\cite{Correa:2018fgz}. Having this one-loop solution, we use it as a starting point for the  iterative procedure~\cite{Gromov:2015vua} 
to obtain a $4$-loop analytico-numerical solution of the QSC (keeping $~200$ digits precision). This is then fed into the purely numerical algorithm of~\cite{Gromov:2015wca} with the new implementation of the optimisation problem. In this way we obtain high precision (around $20$ digits) data for the spectrum (see Fig.~\ref{fig:spectrum}).
The fact that those initial points lead to a convergent procedure for the QSC at finite $g$ is a very nontrivial test of the above construction.

In addition to the scalar sector, other states can, for example, include  covariant derivatives ${\cal D}_t$ and also some combinations of fermions.

The simplest way to describe those states at one loop is again by means of the effective ``doubling-trick"  oscillator numbers 
$\big[n_{\mathbf{a}_1},n_{\mathbf{a}_2}|n_{\mathbf{f}_1},n_{\mathbf{f}_2},n_{\mathbf{f}_3},n_{\mathbf{f}_4}|n_{\mathbf{b}_1},n_{\mathbf{b}_2}\big]$, which in general have to be set to
\begin{align}\la{onum}
\big[\Delta_0-T,\Delta_0-T|1+T,1+T, T, T|\Delta_0-T,\Delta_0-T\big]\;.
\end{align}
where $\Delta_0-T$, roughly, corresponds to the number of covariant derivatives ${\cal D}_t$ (which can also potentially mix with fermions even at one loop).
Note that, whereas $T$ changes the Bethe roots numbers, it does not affect the quantum numbers of the states.
One can call the parameter $T$ a twist, in analogy with higher dimensional cases.
Above, $T=2,\dots,\Delta_0$, except for $\Delta_0=1$ where $T=1$ (this state is exceptional as it satisfies a  semi-shortening condition at weak coupling).
The oscillator numbers~\eq{onum} then lead to a set of $1$-loop
states, which  are used for numerical calculation as before.

At weak coupling, for $\Delta_0 = 1,2,3$ our procedure produces $1,2,6$ solutions respectively, and for $\Delta_0=4$ we computed $19$ levels  (which are possibly not an exhaustive list). 
From $\Delta_0 = 5$ onward, we only solved for twist-$2$ states, which dominate at weak coupling.
All those states proceed to constant integers $\Delta_\infty$
at strong coupling, in contrast with the early expectations~\cite{Alday_2007} and also very different to the  spectrum of local operators, where all unprotected states scale to infinity.
For $\Delta_\infty = 2,4,6,7,8$, we found $1,2,4,1,9$ states respectively, which matches with the counting of~\cite{carlotalk}.
As we only computed $35$ states, at higher $\Delta_\infty$, we already miss some levels  (for example for $\Delta_\infty =9$ we are $2$ states short).

\section{4pt Function and Crossing symmetry}

In order to extract the structure constant $C_1$
we consider the correlator of $4$ arbitrary operators belonging to the contour deformation multiplet ${\cal B}_1$. All such $4$-point functions are related and can be expressed in terms of a single nontrivial function $f(\chi)$~\cite{Liendo:2016ymz,Liendo:2018ukf,Ferrero:2021bsb} thanks to the analytic  superspace formalism of~\cite{Howe:2001je}. For example for four identical scalars:
\beqa\la{pt4}
   && \langle \langle \Phi_{\perp}^1(x_1) \Phi_{\perp}^1(x_2) \Phi_{\perp}^1(x_3) \Phi_{\perp}^1(x_4)\rangle\rangle =\\
\nonumber    &&\frac{ F \chi^2 +  (2 \chi^{-1} - 1)f -\left(\chi^2 - \chi +1\right)f' }{x_{12}^2  \, x_{34}^2 }
\eeqa
where $\chi=\frac{x_{12} x_{34}}{x_{13} x_{24}}$, $x_{ij} = x_i-x_j$.
Finally, $F = 1 + C^2_{\rm BPS}=
\frac{3 W W''}{(W')^2}$ with $W=\frac{2 I_1(\sqrt{\lambda })}{\sqrt{\lambda }}$ \cite{Erickson:2000af,Drukker:2000rr,Pestun:2009nn}. 
The function \eq{pt4} has a symmetry under
the cyclic permutation of the coordinates $x_i$, which translates into the crossing equation:
\beq\la{crossing}
(1-\chi)^2 f(\chi) + \chi^2 f(1-\chi)=0\;.
\eeq
Furthermore, since $\Phi^1_{\perp}$ is a super-conformal primary $f(\chi)$ can be decomposed into a sum over conformal blocks
\begin{equation}
f(\chi) = \chi + { {C^2_{\rm BPS} \,  {F}_{\mathcal{B}_2}(\chi)}}  + \sum_{n } { {C^2_{n} \,  {F}_{{\Delta_n}}(\chi)}}
\end{equation}
where $F_{\mathcal{B}_2} = \chi - \chi\, _2F_1(1,2,4;\chi ) $ and
\beqa\la{Gcrossing}
F_{{\Delta}} &=& \frac{\chi^{\Delta+1}}{1-\Delta}\, _2F_1(\Delta+1,\Delta+2,2 \Delta+4;\chi )\;.
\eeqa
It is clear that the equation \eq{crossing}
can be written as
\beq\la{linC}
\sum_{n}C_n^2 G_{\Delta_n}(\chi) = H(\chi)\;,
\eeq
where the functions $G$ and $H$ are known.
Assuming that we have access to the full spectrum,~\eq{linC} becomes a system of linear equations for $C_n^2$. An obvious problem is that
\eq{linC} contains an infinite number of equations (by picking various values of $\chi$) for an infinite number of unknowns $\{C_n^2\}_{n=1}^\infty$. 
As we managed to compute a large number of low lying states $\Delta_n$ from integrability,
the main hurdle is   finding an efficient truncation scheme which would allow one to obtain a good approximation (at least numerically) for the OPE coefficients $C_n^2$. In the next section we describe several possibilities and  our main result for $C^2_1$ given in Fig.~\ref{fig:C1plot}.

\section{Solving the crossing equation}
In this section we report on various attempts to truncate 
the crossing equation~\eq{Gcrossing} to a finite dimensional system.
\paragraph{Point-like functionals.} 
One of the obvious ways to get a finite system out of~\eq{Gcrossing} is to truncate the sum at some level $\Delta_N$ and sample $N$ different values of the cross-ratio $\chi_i$.
One can then solve a $N\times N$ linear system for the OPE coefficients $a_n\equiv C_n^2$. Assuming the procedure converges, different sets of points $\{\chi_i\}$ should give similar results, but in practice that is not the case.
It was proposed in  \cite{Picco:2016ilr} to average the result over a large number of sampling sets, and use the statistical variance of $a_n$ as an error estimate. Even though this method at the first sight is quite unusual, it does give good results in situations where the density of the spectrum does not increase too fast with $\Delta$. Indeed it worked well in the context of the 2D critical lattice models studied in \cite{Picco:2016ilr,He:2020rfk}.
In our case, the number of states increases rapidly and this method gives almost $100\%$ error in the intermediate coupling region.  Similar problems were observed also by the authors of  \cite{He:2020rfk}. 

We attempted to improve the method slightly by making it more deterministic. First we rewrite the truncated equation \eq{linC} as
$\sum_{n=0}^N a_n G_n(\chi)=0$ with $G_0=H$ and $a_0=-1$. Then we sample $M>N$ points $\chi_i$ and find $a_n$ by minimisation of 
\beq
S(\{a_n\})=\sum_{i=1}^M\[\sum_{n=0}^N a_n G_n(\chi_i)\]^2\;.
\eeq
This method converges  better, but still gives large error bars in our case. Again, for a spectrum whose density does not grow too fast it works well, e.g. for the free spectrum $\Delta_n=2n$ it gives for $N=6$ the known result $a_1 = \frac{2}{5}$
 with  $7$ digits. This number also gives the strong coupling asymptote of our result in  Fig.~\ref{fig:C1plot}.

\paragraph{Oscillating optimised functionals.}
In a more abstract way one can define a linear functional 
$\alpha$ which, when acting on  $G_{\Delta_n}(\chi)$, returns a number $\omega(\Delta_n)$. For the case of the point-like functional, $\alpha$ simply evaluates its argument as some point $\chi_i$.
In order for the procedure of extracting the OPE coefficients to work well, one should make sure that $\omega(\Delta)$ is decaying fast at large $\Delta$. 
Whereas the point-like functional does decay with $\Delta$, one can construct much faster decaying functionals. A simple way is to look for such functionals in the form
\beq\la{cns}
\alpha[f]=\sum_{n=0}^{N/2} \left.c_n \d^{2n} f(\chi)\right|_{\chi=1/2} 
\eeq
for some large $N$. One can find the coefficients $c_n$
from the requirement that the corresponding $\omega(\Delta)$ is small (w.r.t. to a suitable measure) for all $\Delta>\Delta_g$. This type of functionals allows one to truncate the system very efficiently and give a result for $C_1^2$, consistent with more traditional optimal positive functionals, which we describe below. The advantage of this method is that it does not require positivity of the OPE coefficients $C_n^2$ and thus could be used for non-unitary theories. The disadvantage is that estimating the error of the approximation requires extra effort, whereas the positive functionals give  exclusion domains from which one can estimate the error immediately. 

\paragraph{Optimal positive functionals.}
\begin{figure}
    \centering
\includegraphics[scale=0.88]{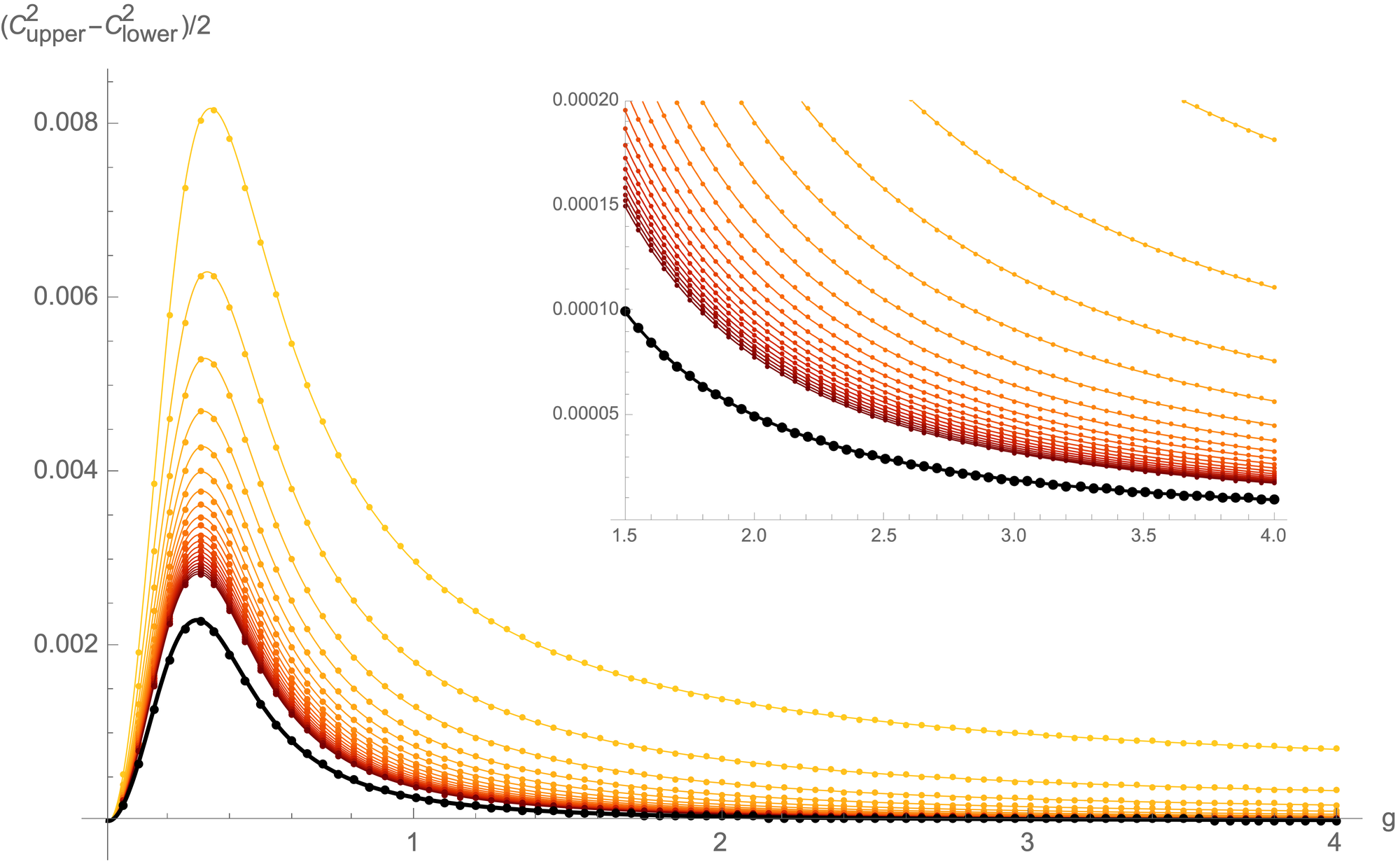}
    \caption{The orange lines show half of the difference between upper and lower bound for $N=5,7,\dots,45$ number of nontrivial derivatives. The black line is an  extrapolation to an infinite number of derivatives. Thus, it gives our estimate for the upper limit of the error of our result.}
    \label{fig:Cgap}
\end{figure}
A more standard  NCB approach is to use optimised functionals which are positive above $\Delta>\Delta_g$.
The SDPB package~\cite{Simmons-Duffin:2015qma,Landry:2019qug} allows one to find such functionals easily. For example, SDPB can find $c_n$ from \eq{cns} such that $\omega(\Delta)>0$ for $\Delta>\Delta_g$, $\omega(\Delta_1)=\pm 1$ and $\alpha[H]$ is minimal. It is easy to see that then $\pm \alpha[H]$ gives upper/lower bound for $C_1^2$ if there are no states in the interval $(\Delta_1,\Delta_{g})$. This methods works very well in our case, by taking $\Delta_g=\Delta_2$ we get a very narrow allowed interval for $C_1^2$ (Fig.~\ref{fig:C1plot}). Half of the length of this interval gives the error of our result as shown on Fig.~\ref{fig:Cgap}. The precision of this method increases with the number of terms in the sum \eq{cns}. By computing the bounds for $N=5,7,\dots,45$ we found that $1/N$ fit give a very stable prediction for $N=\infty$ limit, with $\sim 10^{-7}$ extrapolation error, which is also shown on Fig.~\ref{fig:Cgap} by a thick black line. This gives our final estimate for the error of our result for $C_1^2$. We see that the estimation for $C_1^2$ works the best above $g>1$, but even at smaller coupling the absolute error is $\lessapprox 0.0023$, it decreases quickly to $\lessapprox 0.0001$ for $g=1.5$ and for $g=4$ it reaches $\lessapprox 0.00001$. The full result is given in Tab.~\ref{tab:bounds}.

\paragraph{Incorporating higher states.}
\begin{figure}[t]
{    \centering
\includegraphics[scale=0.88]{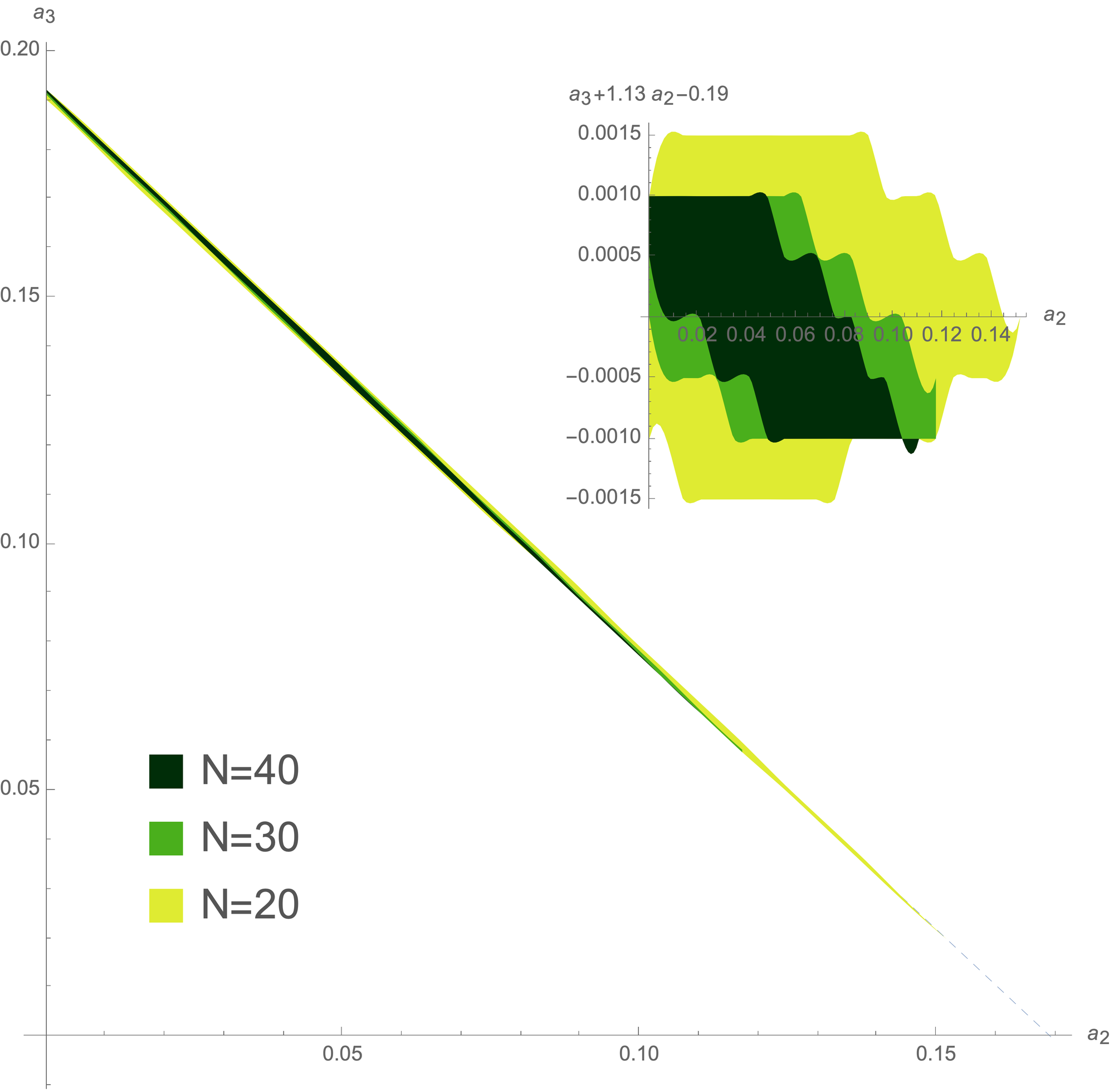}
}
    \caption{\small Bounds for OPE coefficients  $a_2=C_2^2,\;a_3=C_3^2$ of the two higher states of bare dimension $2$ for $g=1$. The very narrow allowed domain is situated along the line $a_3+1.13 a_2=0.19$ (dashed line), which shrinks further with increased $N$.
    }
    \label{fig:C2C3plot}
\end{figure}
What is very striking is that the NCB already gives very precise bounds given only two states $\Delta_1$ and $\Delta_2$ as an input. In order to improve it further one can consider more states and different types of optimisation problems. The numerical analysis with current methods becomes more complicated. Here we present some preliminary results, which incorporate two more states.

In Fig.~\ref{fig:C2C3plot} we plot the allowed regions for $C_2$ and $C_3$. The NCB gives a very narrow domain, which shrinks further with increased $N$.
The reason the domain is stretched in one direction is  the fact that these two states are situated rather close to each other, and it requires higher resolution and large $N$ to resolve between them. Notably, for each allowed value of $C_2$ and $C_3$ the corresponding allowed region for $C_1$ is much  narrower than we found previously. This gives a hope that the bounds on $C_1$ could be further improved. It remains, however, an open question if one can reach an arbitrary high precision for $C_1$, or if there is a fundamental limit. 
We reserve these questions  for  future study.

\section{Conclusions}
The combination of integrability and conformal bootstrap methods gave us surprisingly precise results. Yet we only used a small amount of data available from integrability. In addition to obvious things such as incorporating more states and increasing numerical precision, we believe there is more information from the integrability side which can be incorporated in the current setup~\cite{upcoming}. Another direction would be to try to use the similar approach of combining QSC with NCB in order to extract some bulk CFT data. Crossing equations in 4D can provide more constraints than the 1D crossing we use here. 

The type of problem we considered here may require some further development on the NCB side. For example, when the spectrum is partially known one can relax the positivity requirement in the intervals between the states, which would potentially give tighter bounds on the OPE coefficients.

In conclusion, whereas we cannot yet claim with certainty that
\begin{center}
\verb"QSC + conformal bootstrap = solution of SYM"
\end{center}
we have produced clear evidence that these two methods work well together, giving us rich insights about the non-perturbative regime of planar ${\cal N}=4$ SYM.

\begin{acknowledgments}
We thank P.~Ferrero, Y.~He, A.~Stergiou, D.~Volin and especially C.~Meneghelli for insightful discussions.
JJ is grateful to M.~Tr\'epanier for helpful discussions.
JJ acknowledges the computing resources at Morphing Machines and the CAD Lab at the Indian Institute of Science which were used to obtain some of the numerical results, and thanks H.~Chambeti, S.~K.~Nandy and R.~Narayan for the same. 
MP thanks A.~Manenti for useful discussions.
The work of AC, NG and MP is supported by European Research Council (ERC) under the European Union’s Horizon 2020 research and innovation programme (grant agreement No. 865075)  EXACTC.
NG is also partially supported by the STFC grant (ST/P000258/1).

\end{acknowledgments}

\bibliography{references}
\newpage
\onecolumngrid
\section*{Supplementary material}

\begin{table*}[h]
\caption{\label{tab:states} 
Below we list  the 35 states which are shown in figure~\ref{fig:spectrum} in the main text. They are classified according to their engineering dimension $\Delta_0$ and twist $T$, which dictates their oscillator content. 
In order to initialise the QSC numerics, we obtained the anomalous dimension of each of these states up to four loops. We display the one-loop anomalous dimensions here. 
}
\begin{ruledtabular}
\begin{tabular}{ccccc}
 state \# & $\Delta_0$ & $T$ & oscillator content & one-loop anomalous dimension ($\times g^2$)
 \\ 
 \midrule[0.08em]
 1 & 1 & 1 & \big[0,0$|$2,2,1,1$|$0,0\big] & 4 \\ \midrule[0.08em] 
 2 & 2 & 2 & \big[0,0$|$3,3,2,2$|$0,0\big] & $5 - \sqrt{5}$ \\
 3 &  &   &  & $5 + \sqrt{5}$ \\  \midrule[0.08em] 
 4 & 3 & 2 & \big[1,1$|$3,3,2,2$|$1,1\big] & $\frac{1}{3}(23-\sqrt{37})$ \\
 5 &  &   &  & $\frac{1}{3}(23+\sqrt{37})$ \\ \midrule[0.02em] 
 6 & 3 & 3 & \big[0,0$|$4,4,3,3$|$0,0\big] & 2.1320074084792687244 \\
 7 &  &   &  & 5.1030837048635124382 \\
 8 &  &   &  & 9 \\
 9 &  &   &  & 11.764908886657218837 \\ \midrule[0.08em] 
 10 & 4 & 2 & \big[2,2$|$3,3,2,2$|$2,2\big] & 4.3827667916296171410 \\
 11 &  &   &  & 8.3592314419863375352 \\
 12 &  &   &  & 11.591335099717378657 \\ \midrule[0.02em] 
 13 & 4 & 3 & \big[1,1$|$4,4,3,3$|$1,1\big] & 3.5849116486623305086 \\
 14 &  &   &  & 6.6465020992983359704 \\
 15 &  &   &  & 7.8452994616207484709 \\ 
 16 &  &   &  & 10.154700538379251529 \\
 17 &  &   &  & 13.164880266242322235 \\
 18 &  &   &  & 13.270372652463677951 \\ \midrule[0.02em] 
 19 & 4 & 4 & \big[0,0$|$5,5,4,4$|$0,0\big] & 1.7415869587206441192 \\
 20 &   &   &  & 3.8207798089291592961 \\
 21 &   &   &  & 4.7681120552522578769 \\
 22 &   &   &  & 6.7639320225002103035 \\
 23 &   &   &  & 8.3957770036278304627 \\
 24 &   &   &  & 9.2246674891658188293 \\
 25 &   &   &  & 11.236067977499789696 \\ 
 26 &   &   &  & 11.589746114145456005 \\
 27 &   &   &  & 14.269863799438536581 \\
 28 &   &   &  & 16.189466770720296828 \\ \midrule[0.08em] 
 29 & 5 & 2 & \big[3,3$|$3,3,2,2$|$3,3\big] & 6.7657787405986762795 \\
 30 &   &   &  & 10.581504245500595412 \\
 31 &   &   &  & 13.119383680567394975 \\ \midrule[0.08em] 
 32 & 6 & 2 & \big[4,4$|$3,3,2,2$|$4,4\big] & 5.5267369134729247383 \\
 33 &   &   &  & 9.1989048906866331090 \\
 34 &   &   &  & 12.342918072928388122 \\
 35 &   &   &  & 14.398106789578720696
 \\
\end{tabular}
\end{ruledtabular}
\end{table*}

\begin{table*}[h]
\caption{\label{tab:bounds} 
Below we present our numerical bounds for the square of the OPE coefficient $C_1$ for a wide range of coupling $g\in [0,4]$. The format of the data for $C_1^2$ is $\bigg[\left(C^2_{\rm upper} + C^2_{\rm lower}\right)/2 \pm \left(C^2_{\rm upper} - C^2_{\rm lower}\right)/2\bigg]$.
}
\begin{ruledtabular}
\begin{tabular}{cc|cc|cc|cc}
 $g$ & $C^2_1$ & $g$ & $C^2_1$ & $g$ & $C^2_1$ & $g$ & $C^2_1$ \\ 
 \midrule[0.08em]
 $0.05$ & $0.00504\pm$ $0.00018$ & $1.05$ & $0.29853\pm 0.00024$ & $2.05$ & $0.34604\pm 0.00005$ & $3.05$ & $0.363312\pm 0.000018$ \\
 $0.10 $ & $ 0.0193\pm 0.0007 $ & $ 1.10 $ & $ 0.30279\pm 0.00021 $ & $ 2.10 $ & $ 0.34728\pm 0.00004 $ & $ 3.10 $ & $ 0.363891\pm 0.000018 $ \\
 $ 0.15 $ & $ 0.0406\pm 0.0013 $ & $ 1.15 $ & $ 0.30671\pm 0.00019 $ & $ 2.15 $ & $ 0.34847\pm 0.00004 $ & $ 3.15 $ & $ 0.364452\pm 0.000017 $ \\
 $ 0.20 $ & $ 0.0663\pm 0.0019 $ & $ 1.20 $ & $ 0.31033\pm 0.00017 $ & $ 2.20 $ & $ 0.34960\pm 0.00004 $ & $ 3.20 $ & $ 0.364995\pm 0.000016 $ \\
$ 0.25 $ & $ 0.0936\pm 0.0022 $ & $ 1.25 $ & $ 0.31368\pm 0.00016 $ & $ 2.25 $ & $ 0.35068\pm 0.00004 $ & $ 3.25 $ & $ 0.365523\pm 0.000016 $ \\ $
 0.30 $ & $ 0.1206\pm 0.0023 $ & $ 1.30 $ & $ 0.31679\pm 0.00014 $ & $ 2.30 $ & $ 0.35172\pm 0.00004 $ & $ 3.30 $ & $ 0.366034\pm 0.000015 $ \\ $
 0.35 $ & $ 0.1458\pm 0.0022 $ & $ 1.35 $ & $ 0.31969\pm 0.00013 $ & $ 2.35 $ & $ 0.352714\pm 0.000034 $ & $ 3.35 $ & $ 0.366531\pm 0.000015 $ \\ $
 0.40 $ & $ 0.1684\pm 0.0019 $ & $ 1.40 $ & $ 0.32239\pm 0.00012 $ & $ 2.40 $ & $ 0.353669\pm 0.000032 $ & $ 3.40 $ & $ 0.367013\pm 0.000014 $ \\ $
 0.45 $ & $ 0.1882\pm 0.0016 $ & $ 1.45 $ & $ 0.32491\pm 0.00011 $ & $ 2.45 $ & $ 0.354586\pm 0.000031 $ & $ 3.45 $ & $ 0.367482\pm 0.000014 $ \\ $
 0.50 $ & $ 0.2053\pm 0.0013 $ & $ 1.50 $ & $ 0.32728\pm 0.00010 $ & $ 2.50 $ & $ 0.355468\pm 0.000029 $ & $ 3.50 $ & $ 0.367938\pm 0.000013 $ \\ 
 $0.55 $ & $ 0.2201\pm 0.0011 $ & $ 1.55 $ & $ 0.32951\pm 0.00009 $ & $ 2.55 $ & $ 0.356317\pm 0.000028 $ & $ 3.55 $ & $ 0.368381\pm 0.000013 $ \\ $
 0.60 $ & $ 0.2329\pm 0.0009 $ & $ 1.60 $ & $ 0.33160\pm 0.00009 $ & $ 2.60 $ & $ 0.357134\pm 0.000027 $ & $ 3.60 $ & $ 0.368812\pm 0.000012 $ \\ $
 0.65 $ & $ 0.2440\pm 0.0008 $ & $ 1.65 $ & $ 0.33357\pm 0.00008 $ & $ 2.65 $ & $ 0.357921\pm 0.000026 $ & $ 3.65 $ & $ 0.369232\pm 0.000012 $ \\ $
 0.70 $ & $ 0.2538\pm 0.0006 $ & $ 1.70 $ & $ 0.33544\pm 0.00007 $ & $ 2.70 $ & $ 0.358679\pm 0.000024 $ & $ 3.70 $ & $ 0.369640\pm 0.000012 $ \\ $
 0.75 $ & $ 0.2624\pm 0.0005 $ & $ 1.75 $ & $ 0.33720\pm 0.00007 $ & $ 2.75 $ & $ 0.359411\pm 0.000023 $ & $ 3.75 $ & $ 0.370038\pm 0.000011 $ \\ $
 0.80 $ & $ 0.2701\pm 0.0005 $ & $ 1.80 $ & $ 0.33887\pm 0.00006 $ & $ 2.80 $ & $ 0.360118\pm 0.000022 $ & $ 3.80 $ & $ 0.370425\pm 0.000011 $ \\ $
 0.85 $ & $ 0.2770\pm 0.0004 $ & $ 1.85 $ & $ 0.34045\pm 0.00006 $ & $ 2.85 $ & $ 0.360800\pm 0.000022 $ & $ 3.85 $ & $ 0.370803\pm 0.000011 $ \\ $
 0.90 $ & $ 0.28318\pm 0.00035 $ & $ 1.90 $ & $ 0.34195\pm 0.00006 $ & $ 2.90 $ & $ 0.361460\pm 0.000021 $ & $ 3.90 $ & $ 0.371171\pm 0.000010 $ \\ $
 0.95 $ & $ 0.28879\pm 0.00030 $ & $ 1.95 $ & $ 0.34338\pm 0.00005 $ & $ 2.95 $ & $ 0.362098\pm 0.000020 $ & $ 3.95 $ & $ 0.371530\pm 0.000010 $ \\ $
 1.00 $ & $ 0.29388\pm 0.00027 $ & $ 2.00 $ & $ 0.34475\pm 0.00005 $ & $ 3.00 $ & $ 0.362715\pm 0.000019 $ & $ 4.00 $ & $ 0.371880\pm 0.000010 $ \\ 
\end{tabular}
\end{ruledtabular}
\end{table*}

\end{document}